\documentclass{desyproc}
\usepackage{paralist}

\begin{document}
\title{Status report of the CERN light shining through the wall experiment with microwave axions and related aspects}

\author{{\slshape M. Betz$^1$, F. Caspers$^1$, M. Gasior$^1$, M. Thumm$^{2}$}\\[1ex]
$^1$European Organization for Nuclear Research (CERN), Geneve, Switzerland\\
$^2$Karlsruhe institute of technology (KIT), Germany}

\contribID{Betz\_Michael}

\desyproc{DESY-PROC-2011-04}
\acronym{Patras 2011} 
\doi  
\linespread{1.0}
\maketitle

\begin{abstract}
One way to proof or exclude the existence of axion like particles is a microwave light shining through the wall experiment. In this publication we will emphasize on the engineering aspects of such a setup, currently under development at CERN. One critical point, to achieve meaningful results, is the electromagnetic shielding between axion-emitter and –receiver cavity, which needs to be in the order of 300 dB to improve over existing experimental bounds. The RF leakage or electromagnetic crosstalk between both cavities must be well controlled and quantified during the complete duration of the experiment. 
A very narrow band (in the $\mathrm{\mu Hz}$ range) homodyne detection method is used to reveal the axion signal from background thermal noise.
The current status of the experiment is presented.
\end{abstract}

\section{Motivation and introduction}
The axion is a hypothetical elementary particle, which emerged originally from a proposal by Peccei
and Quinn, intended to solve the strong CP problem \cite{src:theory1}.
The axion is neutral, only interacts very weakly with matter, has very low mass ($\approx 10^{-4} eV/c^2$), spin
zero, and a natural decay constant (to 2 photons) in the order of $10^{17}$ years.

One way to detect axion like particles (ALPs) is to exploit their property of mixing with photons inside a static electric or magnetic field. In a simplified way, this provides ``virtual photons'', forcing ALPs to decay 
into real photons which then can be detected conventionally \cite{src:mixing}. All axion searches nowadays are based on this so called Primakoff effect. 

\begin{figure}[hb]
\centerline{\includegraphics[width=0.9\textwidth]{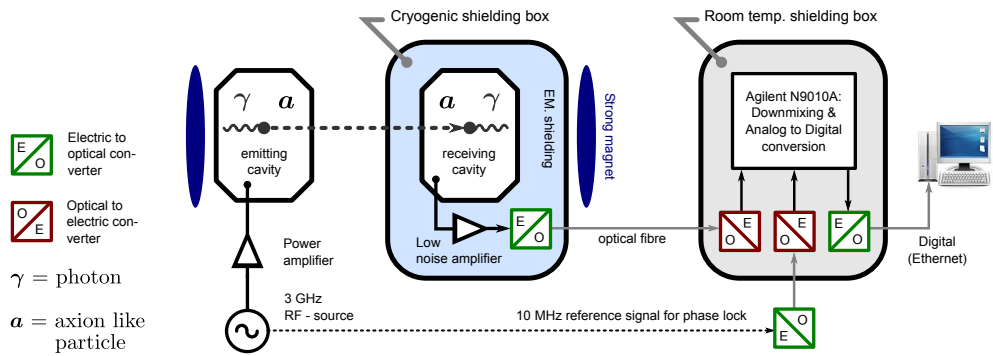}}
\caption{Overview of the light shining through the wall experiment in the microwave range.}
\label{fig:overView}
\end{figure}

A particular sensitive experiment is the light-shining-through-wall (LSW) setup, depicted in Fig.~\ref{fig:overView}. The emitting part consists of a photon source (e.g., a microwave, laser or X-ray beam) and a strong magnet. By the Primakoff effect some photons convert to ALPs. A wall blocks the unconverted photons but not the ALPs, which traverse to the receiving part. There the reciprocal conversion takes place. The ``reconverted'' photons can be detected by conventional means and will have the same wavelength as the photons in the emitting part, considering that no energy is lost at any point. It shall be pointed out that a LSW experiment is also sensitive to hidden photons, which are a different kind of hypothetical particle \cite{src:JaCaRi}. The only necessary modification is not to use an external magnetic field.

The sensitivity of the experiment can be significantly increased by placing resonating structures in the emitting and detection part. Resonator cavities with particular low loss can be constructed in the microwave regime. Q-factors $> 10^4$ can routinely be achieved with normal conducting cavities at room temperature. Q-factors of $\approx 10^{10}$ are technically feasible with superconducting cavities (however, this technology is not compatible with strong magnetic fields). Another advantage is that sensitive homodyne detection methods can be applied in the microwave range, further increasing sensitivity.

For the aforementioned reasons and because of the fact that the instruments, tools and facilities for building microwave cavities for particle accelerators already exist at CERN, a microwaves shining through the wall experiment will be setup in the course of the next 2 years.

\section{Cavity design and construction}
The operating frequency of the emitting cavity determines the energy of the photons inside ($E_{\mathrm{ph}} = h \cdot f$). If the Primakoff effect takes place, the energy of the photon is converted to mass and kinetic energy ($\beta = v / c$) of the ALP as shown in Equ.~\ref{equ:a0}. Therefore, as a first approximation, the choice of frequency determines an upper limit on ALP mass ($m_a$). As a rule of thumb $m_a = 4.14~ \mathrm{\mu eV/c^2}$ corresponds to $f = 1$ GHz.
\begin{align}
\label{equ:a0}
 E_{\mathrm{ph}} = h f = E_{\mathrm{ALP}} = m_a c^2 (1 + \frac{1}{2} \beta^2)
\end{align}
For the first experimental phase, operating at room temperature, testing the electromagnetic (EM) shielding and achieving a leak tight setup is of primary importance. For this phase, the $\mathrm{TE}_{011}$ mode will be used, as it has the highest Q-factor compared to other modes. Hidden photon search without a magnet is possible, however, this mode is not sensitive to ALPs in a magnetic field as its geometric overlap integral \cite{src:hoo, src:ring} is zero. ALPs will be searched in the second phase of the experiment with the fundamental $\mathrm{TM}_{010}$ mode at 1.75 GHz. Higher order modes can be used for ALPs search as well, however certain requirements regarding the magnet and cavity design need to be fulfilled \cite{src:directSearch,src:wiggler}. 

The shape of the cavity used in this test setup is a classical pillbox with bevelled edges. The bevelling helps to separate the degeneracy of the $\mathrm{TM}_{111}$ and the $\mathrm{TE}_{011}$ mode.

To minimize losses, the surface of the brass cavities is coated with a $10~\mu m$ thick layer of silver. A $< 0.2~\mu m$ thick layer of gold prevents oxidation. The skin depth in silver at 3 GHz at room temperature is $\approx 1~\mu m$, so most of the surface current flows in the silver layer with low losses \cite{src:EMI}.

Power is transferred towards and out of the cavity by an inductive coupling loop. By adjusting the angle of the loop's cross section, the cavity impedance can be matched to a $50~\Omega$ system, providing maximum power transfer.

To compensate manufacturing tolerances, both cavities are equipped with a fine threaded tuning bolt, directly perturbing the cavity volume.  The 20 mm diameter bolt provides $\approx 10$ MHz adjustment range of the resonant frequency. For the time being, no sweep over the ALP mass during an experimental run is foreseen. Once the cavities have reached thermal equilibrium, their resonant frequency is sufficiently stable and no manual adjustment of the tune is needed during an experimental run. The resonant frequency of the emitting cavity is continuously monitored by means of its reflected power. The tune of the (room temperature) receiving cavity is determined from the recorded data by means of evaluating the spectral noise power density.   

\section{EMI shielding concept}
There are three categories of perturbation which may disguise the useful signal from converted ALPs and which need to be mitigated:
\begin{compactitem}
  \item Environmental interference such as signals from cellphones, WIFI devices, radio stations, etc.
  \item Direct microwave leakage from the emitting cavity (no photon-ALP-photon conversion).
  \item Thermal noise, which is the random movement of charge carriers in a conductor at non-zero temperature.
\end{compactitem}
The first two points can be diminished by electromagnetic shielding, enclosing the receiving cavity and the signal processing electronics. To minimize the third point, the receiving cavity and the first amplifier need to be cooled to cryogenic temperatures.

We require the screening attenuation of a shielding enclosure to be $\approx 100$ dB (that means unwanted signals are attenuated by a factor of $10^{10}$ in power). Stacking several enclosures (the cavity and cryostat walls count as shielding too) we can reach a total screening attenuation of 300 dB, which is needed to achieve the desired sensitivity and realize reliable detection of an ALP signal with $\mathrm{P_{signal} \approx -230 dBm}$. 

As the detection cavity and first amplifier will need to be placed in a strong magnetic field inside a cryogenic environment (in the second planning phase), the experiment is split into two parts, equipped with two separate shielding enclosures. The cryogenic part (Fig.~\ref{fig:overView}, middle) contains the receiving cavity and first amplifier. The room temperature part (Fig.~\ref{fig:overView} right) contains the signal processing electronics. Both are connected by an analog optical link with $> 3$ GHz bandwidth. An optical fibre guides the noise like signal from the detection cavity to the measurement instruments, unaffected by ambient EM interference and without comprising the performance of the EM shielding enclosures.

Power is provided to the instruments in the shielding enclosures by a custom made feed-trough filter, blocking EMI signals from the outside. The filter consists of several L and C elements in a $\pi$- or T-type configuration \cite{src:EMI}. An example for a L element is a tight fitting wire inside a ferrite tube, which represents a lossy inductor at microwave frequencies. An example of a C element is a commercial ceramic feed-trough terminal filter, which represents a low impedance short circuit at microwave frequencies.

Some of the practical measures which have been applied to improve robustness against EMI include: placing EM absorbing material between shielding layers, incorporating RF-gaskets on removable lids, applying conductive caulking material at material joints, etc.

\section{Online shielding diagnostics}
If there is a signal observed on the detection side of the experiment, it either originates from converted ALPs or from direct EM coupling as a consequence of microwave leakage. We need reliable ways to distinguish the two cases. 
One possibility is to use a slowly time varying (e.g., 1 cycle per hour) magnetic field. We would be able to detect the amplitude modulation sidebands only for an ALP signal but not for electromagnetic leakage.
Another way is to look for small phase differences. Opposed to photons, ALPs have a non-zero rest mass and thus do not travel with the speed of light. Therefore we expect a small phase offset between the two signals.

Another, more robust and generally applicable way is to monitor and record the level of EM leakage during the experimental run. This can be done by emitting low power (in the $\mathrm{\mu W}$ range) probe signals in the laboratory space and between the shielding layers. Monitoring the strength of a signal after transversing one shielding layer gives information about its effective screening performance. Observing a probe signal with a level higher than a certain threshold indicates an EM leak and serves as a veto, rendering a positive result on ALP detection invalid. Each probe signal is emitted on a slightly different (phase locked) frequency so they can be identified and do not interfere with the detection of ALPs. The probe signals are guided inside the shielding layers by analog fibre-optic links.

\section{Signal processing}
\begin{wrapfigure}{r}{0.5\textwidth}
\centerline{\includegraphics[width=0.5\textwidth]{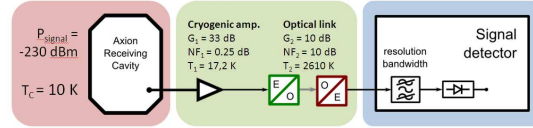}}
\vspace{-3mm}
\caption{Analog signal processing chain.}\label{Fig:asp}
\vspace{-2mm}
\end{wrapfigure}
A conceptual signal processing chain, which might be used for a cryogenic LSW setup, is depicted in Fig.~\ref{Fig:asp}. The smallest signal power at the coupling antenna of the cavity, determined by the specified sensitivity of the experiment is $P_{\mathrm{signal}} = -230$ dBm, corresponding to one 3 GHz photon every 3 minutes.
\begin{align}
\label{equ:NF}
F_n = 10^{\mathrm{NF}_n / 10} \quad \quad \quad F_{equ} = F_1 + \frac{F_2 - 1}{G_1} \quad \quad \quad
T_n = 290 K \cdot \left( F_n - 1\right)
\end{align}
The first amplifier in immediate vicinity to the cavity will be a cryogenic HEMT low noise amplifier, for which rather pessimistic gain and noise temperature values have been assumed in this example. While a typical analog optical link has a noise temperature\footnotemark~of 2610 K, its influence on the overall system is low. The equivalent noise temperature is determined by Friis formula (Equ~\ref{equ:NF}) $T_{equ} = 18.5$ K. The overall system noise temperature is $T'_{\mathrm{noise}} = T_C + T_{equ} = 28.5$ K or in terms of power density at the detector (taking amplifier gain into account) $P'_{\mathrm{noise}} = -141$ dBm / Hz. The signal power at the detector is $P'_{\mathrm{signal}} = -187$ dBm. To upraise the signal from the noise, narrow band filtering is necessary. 

\footnotetext{
We use the generally accepted Rayleigh-Jeans noise equation $P=kTB$, which is an approximation valid for $hf \ll kT$ (h is Plank’s constant, f the frequency, k Boltzmann’s constant, and T the temperature) \cite{src:noise}.
}

The filtering can be done by a commercial Fast Fourier Transform (FFT) type analyzer with the assumption that the photon to ALPs conversion and its reciprocal do not change the spectral properties of the signal and that it is defined with the  spectral resolution of the original RF source. This allows to use very narrow resolution bandwidths ($B_{\mathrm{res}} \leq 25~\mu$Hz), scaling down the noise power in one bin by $P_{bin} = P'_{\mathrm{noise}} \cdot B_{\mathrm{res}}$, nonetheless the signal power in this bin stays constant. To achieve the desired sensitivity with the previous assumptions and a signal to noise ratio of 1 we would need $B_{\mathrm{res}} = 25~\mu$Hz which demands a minimum length of the time trace of $T = 1 / B_{\mathrm{res}} = 11~\mathrm{hours}$.

The RF-source and signal analyzer need to operate on the same frequency (within $B_{\mathrm{res}}$) during the experimental run, otherwise the signal power will smear out over several bins in the spectrum and the signal to noise ratio will degrade. While absolute frequency drifts are unavoidable, phase-locking RF-source and signal analyzer to a common frequency reference allows to achieve a good relative frequency stability. This has been explained in \cite{src:narrowband} and has been successfully demonstrated with resolution bandwidths down to $10~\mu$Hz.

\section{Results of the first measurement-run}
\begin{wrapfigure}{r}{0.6\textwidth}
\centerline{\includegraphics[width=0.6\textwidth]{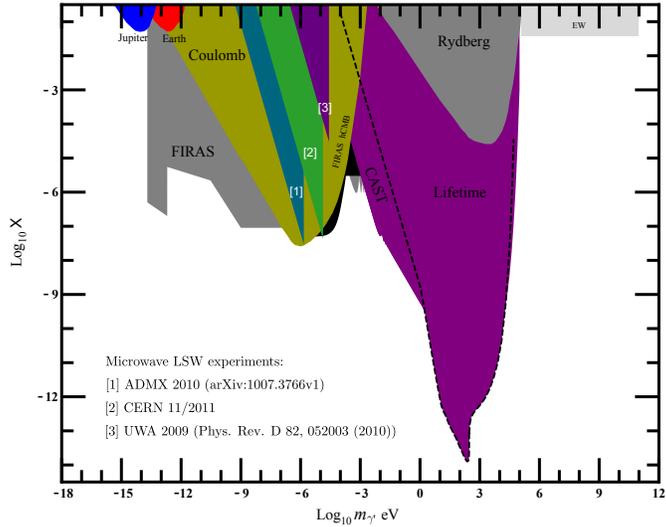}}
\caption{
Preliminary exclusion limit for hidden photons as a result of the non observation of a signal in the measurement-run from the 11/2011 at CERN (labeled as [2]), compared to the result of other experiments. The collection of experimental data has been taken from \cite{src:JaCaRi}, where it is also explained in detail.
}
\label{fig:exclusion}
\end{wrapfigure}
A first measurement-run over 6 h, searching for hidden photons without an external magnetic field at room temperature has been accomplished in  Nov.~2011. The non observation of a signal in the recorded data indicates a new preliminary exclusion limit with a sensitivity to the kinetic mixing parameter down to $\chi = 5 \cdot 10^{-8}$ at a hidden photon mass of $12.23~\mathrm{\mu eV/c^2}$ . This is a slight improvement over the Coulomb and FIRAS experiments, which also have produced exclusion limits in this energy range \cite{src:JaCaRi}. The new (preliminary) limit is depicted in Fig.~\ref{fig:exclusion}, the technical parameters are summarized in Table \ref{tbl:data}. It shall be pointed out, that for the preliminary results, a rather pessimistic assumption for the hidden photon geometric factor of $|G| = 0.01$ has been used. The cavities were operated on the TE$_{011}$ mode and placed next to each other, separated by a distance of 150 mm. For this configuration, we expect the actual value of $|G|$ to be significantly higher than the assumed one.
\begin{table}
\centering
\caption{Technical param. for the hidden photon measurement run from Nov. 2011 at CERN.}
\begin{tabular}{|r|r|l||r|r|l|}
\hline
$f_{em}$			& 2.9571 GHz				& Operating frequency 				&$Q_{\mathrm{em}}$	& 23416						& Q-factor of emitting cav.\\
$P_{\mathrm{em}}$ 	& 49 W 						& Power to emitting cav.			&$Q_{\mathrm{det}}$	& 23620						& Q-factor of detector cav.\\ 
$P_{\mathrm{det}}$	& $1.74 \cdot 10^{-23}$ W	& Detectable sign. power	&$|G|$				& 0.01						& Geometric factor\\
\hline
\end{tabular}
\label{tbl:data}
\end{table}

\section{Conclusion and Outlook}
Mitigating electromagnetic interference is a critical point for sensitive light shining through the wall experiments in the microwave range.
We propose a shielding concept, involving several metal shells and signal transmission over optical fibre. The injection of probe signals between the shielding layers allows to quantify the overall screening attenuation in real time. This allows to make sure that a hypothetical detected signal does not originate from environmental interference.

To receive very weak signals we proposed a narrow band detection scheme. A commercial signal analyser can be operated as homodyne detector by phase locking it to the RF-source driving the emitting cavity. Signals with a power level in the order of $\mathrm{P_{signal} \approx -230~dBm}$ can still be detected.

The first measurement-run on 11/2011 produced new exclusion limits for hidden photons, it showed that the aforementioned concepts are practical and also necessary for a sensitive LSW experiment in the microwave range. 

\section{Acknowledgments}
\vspace{-1mm}
The authors would like to thank R. Jones and the BE department management for encouragement. Thanks to K. Zioutas and the organizers of the Patras Workshop for a very enjoyable and inspiring conference. 
Supported by the Wolfgang-Gentner-Programme of the Bundesministerium f\"ur Bildung und Forschung (BMBF).


\begin{footnotesize}

\end{footnotesize}


\end{document}